\documentclass[twocolumn]{aastex62} 

\usepackage{xspace}
\usepackage{graphicx}
\usepackage{natbib}
\usepackage{amsmath}
\usepackage{booktabs}
\usepackage{multirow}
\usepackage{afterpage}
\usepackage{rotating}

\newcommand{\hide}[1]{}

\newcommand{\y}[1]{\ensuremath{y_{#1}}\xspace}

\newcommand{\gb}{\ensuremath{{\it b}}\xspace}
\newcommand{\absb}{\ensuremath{\vert\,\gb\,\vert}\xspace}

\newcommand{\kms}{\ensuremath{\,{\rm km\,s^{-1}}}\xspace}

\newcommand{\K}{\ensuremath{\,{\rm K}}\xspace}

\newcommand{\ghz}{\ensuremath{\,{\rm GHz}}\xspace}

\newcommand{\degree}{\ensuremath{^\circ}\xspace}

\newcommand{\arcmper}{\ensuremath{\rlap.{^{\prime}}}}

\newcommand{\hii}{{\rm H\,{\footnotesize II}}\xspace}

\newcommand{\he}[1]{\ensuremath{^{#1}{\rm He}}\xspace}

\newcommand{\hna}{\ensuremath{{\rm H}\,{\rm n}\,\alpha\xspace}}

\usepackage{mathtools}

\begin{document}

\title{Methods for Averaging Spectral Line Data}

\author[0000-0001-8800-1793]{L.~D.~Anderson}
\affiliation{Department of Physics and Astronomy, West Virginia University, Morgantown, WV 26506, USA}
\affiliation{Adjunct Astronomer at the Green Bank Observatory, P.O. Box 2, Green Bank, WV 24944}
\affiliation{Center for Gravitational Waves and Cosmology, West Virginia University, Chestnut Ridge Research Building, Morgantown, WV 26505, USA}

\author[0000-0002-1311-8839]{B. Liu}
\affiliation{National Astronomical Observatories, Chinese Academy of Sciences, Beijing 100101, People’s Republic of China}

\author[0000-0002-2465-7803]{Dana~S.~Balser}
\affiliation{National Radio Astronomy Observatory, 520 Edgemont Road, Charlottesville, VA 22903, USA}

\author[0000-0003-4866-460X]{T.~M.~Bania}
\affiliation{Institute for Astrophysical Research, Astronomy Department, Boston University, 725 Commonwealth Ave., Boston, MA 02215, USA}

\author[0000-0002-9947-6396]{L.~M.~Haffner}
\affiliation{Department of Physical Sciences, Embry-Riddle Aeronautical University, Daytona Beach FL 32114, USA}

\author[0000-0002-4727-7619]{Dylan~J.~Linville}
\affiliation{Department of Physics and Astronomy, West Virginia University, Morgantown, WV 26506, USA}
\affiliation{Center for Gravitational Waves and Cosmology, West Virginia University, Chestnut Ridge Research Building, Morgantown, WV 26505, USA}

\author[0000-0001-8061-216X]{Matteo~Luisi}
\affiliation{Center for Gravitational Waves and Cosmology, West Virginia University, Chestnut Ridge Research Building, Morgantown, WV 26505, USA}
\affiliation{Department of Physics, Westminster College, New Wilmington, PA 16172, USA}

\author[0000-0003-0640-7787]{Trey V. Wenger}
\affiliation{NSF Astronomy \& Astrophysics Postdoctoral Fellow,
  Department of Astronomy, University of Wisconsin--Madison,
  Madison, WI, 53706, USA}

\correspondingauthor{L.D.~Anderson}
\email{loren.anderson@mail.wvu.edu}

\begin{abstract}
The ideal spectral averaging method depends on one's science goals and the available information about one's data.  Including low-quality data in the average can decrease the signal-to-noise ratio (SNR), which may necessitate an optimization method or a consideration of different weighting schemes.  Here, we explore a variety of spectral averaging methods.  We investigate the use of three weighting schemes during averaging: weighting by the signal divided by the variance (``intensity-noise weighting''), weighting by the inverse of the variance (``noise weighting''), and uniform weighting.  Whereas for intensity-noise weighting the SNR is maximized when all spectra are averaged, for noise and uniform weighting we find that averaging the 35-45\% of spectra with the highest SNR results in the highest SNR average spectrum.  With this intensity cutoff, the average spectrum with noise or uniform weighting has $\sim 95\%$ of the intensity of the spectrum created from intensity-noise weighting.  We apply our spectral averaging methods to GBT Diffuse Ionized Gas (GDIGS) hydrogen radio recombination line (RRL) data to determine the ionic abundance ratio, $y^+$, and discuss future applications of the methodology.  

\end{abstract}


\section{Introduction} \label{sec:intro}
Averaging spectral line data allows one to increase the
signal-to-noise ratio (SNR) of the resultant spectrum.  Such averaging is straightforward when the
spectra are taken of the same source and have similar noise characteristics.  The situation is complicated, however, if the noise or source intensity differ significantly between observations.  In
such cases, depending on the distribution of peak intensities in the
observations and the weighting scheme, averaging spectra can result in a
decrease in the SNR.  We
are interested in exploring the implications of using one averaging or weighting method over another.

This is not an entirely new problem but general prescriptions are lacking in the astronomical literature. \citet{rosales-ortega12} explored how to maximize the SNR for 
integral field spectroscopy (IFS) observations and provided code to the community. They argued that the optimal integration method depends on the science case.  \citet{zhang99} dealt with the problem of averaging multiple spectra taken from a chromatography/
spectroscopy experiment.  They found that for a Gaussian-peaked signal distribution, the maximum SNR is attained when the 38\% highest SNR individual spectra are averaged. \citet{unser90} developed a method for maximizing the SNR for a set of 2-D images that better accounts for noisy data where the SNR of individual observations is difficult to measure.  Adaptive smoothing of 2D images, such as using Voronoi Tessellations or Weighted Voronoi Tessellations, can be used to create spatial regions that meet user-specified SNR criteria \citep{cappellari03, diehl06}.  The ideal averaging method may depend on whether the intensities of the spectra to be averaged are uniform or have a large variance.

In this paper, we explore methods for spectral averaging and provide guidance for
multiple use-cases.  We focus our analytical treatment on radio
spectroscopic observations (i.e., we use the variable ``$T$'' for intensity), but the method is applicable to any
spectral line data set.

\section{The Signal to Noise Ratio\label{sec:method}}
A spectral line has a SNR given by \citep{lenz92}
\begin{equation}
  {\rm SNR} = C\left(\frac{\Delta V}{\Delta\lambda}\right)^{0.5}\frac{\langle T_P \rangle}{\langle \sigma \rangle} \,,
  \label{eq:snr_lenz}
\end{equation}
where $C$ is a constant whose value depends on the line shape, $\Delta V$ is the full width at half-maximum (FWHM) line
width, $\Delta\lambda$ is the spectral resolution (or width of the smoothing kernel),  $\langle T_{P} \rangle$ is the average peak line brightness temperature, and $\langle \sigma \rangle$ is
the average rms noise.  For a spectrum that can be modeled as a Gaussian line
with white noise, $C = 0.7$.

Upon averaging $n$ spectra each with (unnormalized) weighting $w_i$, the average intensity at a given spectral channel is 
\begin{equation}
  \langle T \rangle = \frac{\sum_{i=1}^n T_{i} w_i}{\sum_{i=1}^n w_i}\,.
  \label{eq:t}
\end{equation}
At the line center, the peak line intensity $\langle T_{P} \rangle$ is therefore
\begin{equation}
  \langle T_{P} \rangle = \frac{\sum_{i=1}^n T_{P,i} w_i}{ \sum_{i=1}^n w_i}\,.
  \label{eq:peak}
\end{equation}
The average (uncorrelated) rms spectral noise is
\begin{equation}
  \langle \sigma \rangle = \left[ \frac{\sum_{i=1}^n \sigma_{i}^2 w_i^2}{\left(\sum_{i=1}^n w_i\right)^2} \right]^{0.5}\,.
    \label{eq:sigma}
\end{equation}
The SNR in the average spectrum is then
\begin{equation}
  {\rm SNR} = C \left(\frac{\Delta V}{\Delta\lambda}\right)^{0.5}\frac{\sum_{i=1}^n T_{P, i} w_i}{\left(\sum_{i=1}^n \sigma_{i}^2 w_i^2\right)^{0.5}}\,.
  \label{eq:snr}
\end{equation}
Equation~\ref{eq:snr} is the fundamental equation that governs the
increase in SNR when averaging multiple spectra with weighting $w_i$, assuming uncorrelated noise.  

We discuss three weighting schemes below.  An observer's choice of weighting is dictated by their science goals and the availability of information about their data.  If the noise and
peak intensity are the same for all spectra such that $\sigma_i = \sigma_0$ and $T_{P,i} = T_{P,0}$, for the weighting schemes considered here Equation~\ref{eq:snr} reduces to
\begin{equation}
  {\rm SNR} = C \left(\frac{\Delta V}{\Delta\lambda}\right)^{0.5} \frac{T_{P,0}}{\sigma_0}n^{0.5}\,.
  \label{eq:snr_const}
\end{equation}
Equation~\ref{eq:snr_const} approximates the SNR when averaging multiple
spectra taken of the same source with the same integration times and observing conditions.
If the noise is correlated between spectra, then the noise term in Equation~\ref{eq:snr} will include covariances between the spectra and the exponent of $n$ in Equation 6 will be less than 0.5. 
One can recover the dependence on $n$ in Equation~\ref{eq:snr_const} by considering only the number of independent spectra.

If the noise is correlated, the average noise decreases slowly when averaging and therefore the exponent in the term $n^{0.5}$ decreases and the SNR increases more slowly than in the uncorrelated case.  One can recover the expected dependence on $n$ by considering the number of independent samples.

\subsection{Intensity-Noise Weighting}
For ``Intensity-Noise Weighting,''
\begin{equation}
  w_i = T_{P, i} \sigma_i^{-2}\,.
    \label{eq:intensity_weighting}
\end{equation}
Using this weighting will bias the average peak line intensity by the highest values of $T_{P,i}$:
\begin{equation}
      {\rm SNR} = C \left(\frac{\Delta V}{\Delta\lambda}\right)^{0.5} \left(\sum_{i=1}^n T_{P, i}^2 \sigma_i^{-2}\right)^{0.5}\,.
      \label{eq:snr_intensity_noise}
\end{equation}
If all spectra have the same (uncorrrelated) noise $\sigma_0$, 
Equation~\ref{eq:snr_intensity_noise} reduces to
\begin{equation}
  {\rm SNR} = C \left(\frac{\Delta V}{\Delta\lambda}\right)^{0.5} \frac{\left(\sum_{i=1}^n T_{P, i}^2\right)^{0.5}}{\sigma_0}\,.
\end{equation}
If all signal strengths $T_{P,0}$ are the same but the noise is variable, as in
the case of averaging data taken of the same source under different observing conditions or integration times, Equation~\ref{eq:snr_intensity_noise} becomes
\begin{equation}
  {\rm SNR} = C \left(\frac{\Delta V}{\Delta\lambda}\right)^{0.5}  T_{P,0} \left( \sum_{i=1}^n \sigma_i^{-2} \right)^{0.5}\,.
    \label{eq:snr_intensity_noise_t_const}
\end{equation}

\subsection{Noise Weighting}
For ``Noise weighting,''    
\begin{equation}
  w_i = \sigma_i^{-2}\,.
\end{equation}
This weighting will bias the average peak line intensity toward spectra with lower noise:
\begin{equation}
  {\rm SNR} = C \left(\frac{\Delta V}{\Delta\lambda}\right)^{0.5} \frac{\sum_{i=1}^n T_{P, i}\sigma^{-2}}{\left(\sum_{i=1}^n \sigma_i^{-2}\right)^{0.5}}\,.
  \label{eq:snr_noise}
\end{equation}
If all spectra have the same (uncorrelated) noise $\sigma_0$,
Equation~\ref{eq:snr_noise} reduces to
\begin{equation}
  {\rm SNR} = C \left(\frac{\Delta V}{\Delta\lambda}\right)^{0.5} \frac{\sum_{i=1}^n T_{P, i}}{\sigma_0} n^{-0.5}\,.
    \label{eq:snr_noise_sigma_const}
\end{equation}
If all signal strengths $T_{P,0}$ are the same, but the noise is variable, we again find Equation~\ref{eq:snr_intensity_noise_t_const}.  

\subsection{Uniform Weighting}
For ``Uniform weighting,''
\begin{equation}
    w_i = 1
\end{equation}
and
\begin{equation}
  {\rm SNR} = C \left(\frac{\Delta V}{\Delta\lambda}\right)^{0.5} \frac{\sum_{i=1}^n T_{P, i}}{(\sum_{i=1}^n \sigma_i^2)^{0.5}}\,.
  \label{eq:snr_uniform}
\end{equation}
If all spectra have the same (uncorrelated) noise $\sigma_0$, we again find Equation~\ref{eq:snr_noise_sigma_const}.
If all signal strengths $T_{P,0}$ are the same but the noise is variable,
we again find Equation~\ref{eq:snr_intensity_noise_t_const}.

\subsection{Maximizing the Signal to Noise Ratio}
For a given weighting scheme and averaging method, the optimal value for $n$ is often found when the SNR reaches a maximum value SNR$_{\rm max}$, or when
\begin{equation}
  \frac{d}{dn} ({\rm SNR}) = 0\,.
\end{equation}
For intensity-noise weighting, or in the case that all spectra have the same values of $T_{P,i}$ and $\sigma_0$, averaging all available spectra will result in the highest SNR (cf. Equation~\ref{eq:snr_const}).  
For noise and uniform weighting, if the values of $T_{P,i}$ or $\sigma_i$ are different, the ideal number of spectra to average may be less than the total number of spectra available.

To determine the ideal number of spectra to average for noise and uniform weighting, our method requires that the spectra be ordered by
decreasing SNR.  For individual spectra,
\begin{equation}
  {\rm SNR}_i \propto \frac{T_{P,i}}{\sigma_i}\,.
  \label{eq:snr_i}
\end{equation}
To determine $n$ and SNR$_{\rm max}$, one therefore must:
\begin{enumerate}
\item compute or estimate the peak line intensity, $T_{P, i}$, and the rms spectral noise, $\sigma_i$, for all spectra;
\item order the spectra in terms of SNR$_i$ (using Equation~\ref{eq:snr_i});
\item determine when the average SNR is maximized (SNR$_{\rm max}$), either theoretically using Equation~\ref{eq:snr} or by fitting the average spectra with a model.
\end{enumerate}
Below, we use this method to estimate SNR$_{\rm max}$ and $n$ for simulated distributions of $T_P$.

\section{Simulated Signal and Noise Distributions}
We perform Monte Carlo simulations
to assess the effects of different intensity and noise distributions, as well as weighting schemes.

\subsection{Distributions for $T_P$}
\begin{figure*}
    \centering
    \includegraphics[width=3in]{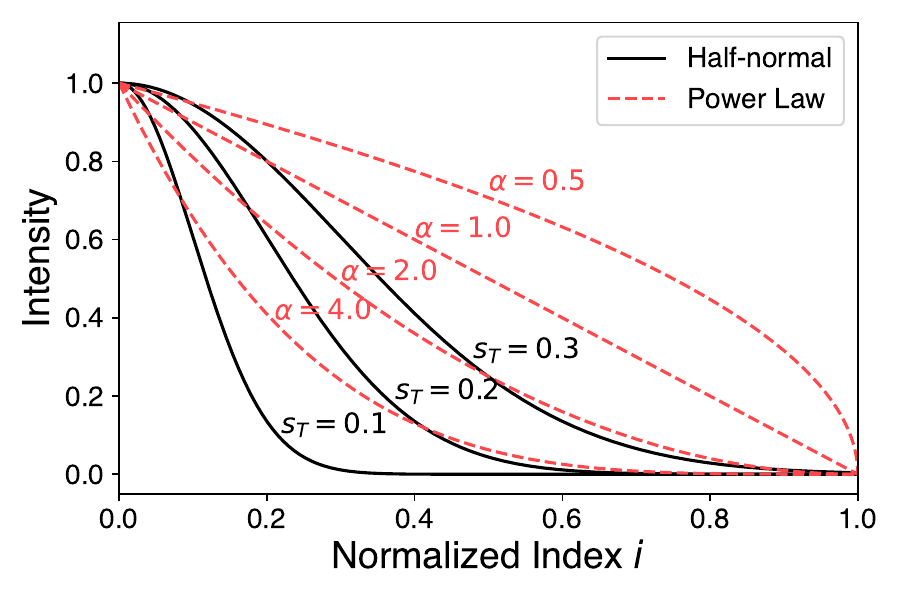}
    \includegraphics[width=3.350in]{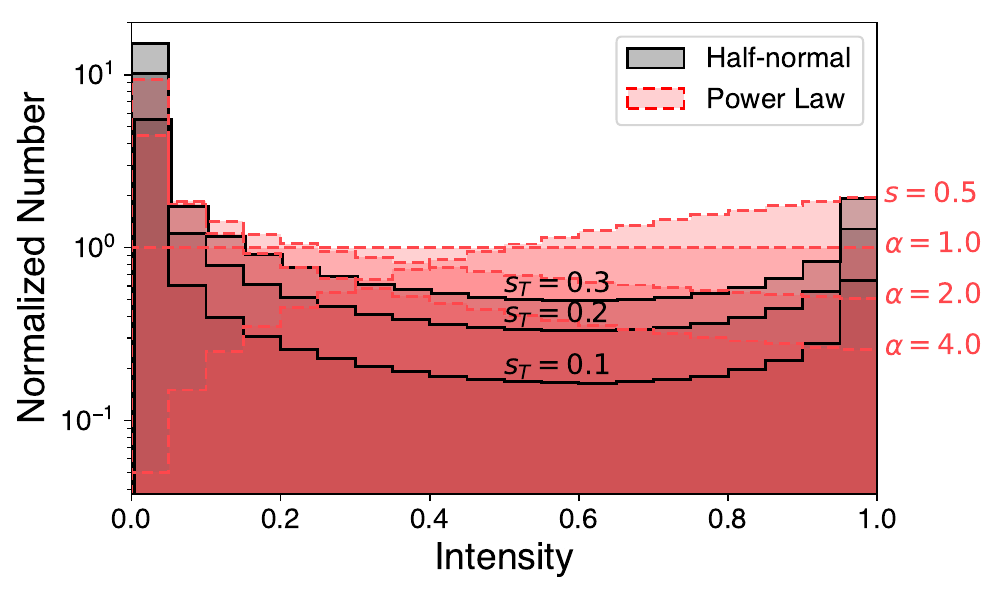}
    \caption{Simulated intensity distributions (see Sections~\ref{sec:gaussian} and \ref{sec:powerlaw}).  The left panel shows the distributions themselves, while the right panel shows a histogram of the values.  Half-normal distributions have a larger number of extremely low intensity values.  Low power law indices have more high intensity values whereas higher power law indices have more low intensity values.}
    \label{fig:simulated}
\end{figure*}

We investigate two characteristic distributions for $T_{P}$: half-normal and power law.  We plot the distributions in  Figure~\ref{fig:simulated}, for a range of half-normal standard deviations (see Section~\ref{sec:gaussian}) and power law indices (see Section~\ref{sec:powerlaw}).  The half-normal distribution is what is measured from a compact source and a Gaussian telescope response, whereas the power law distributions are meant to model diffuse (low power law indices) and compact (high power law indices) sources.  For both distributions, we assume that the distribution of noise values is Gaussian, characterized by a mean value of $\sigma_0$ and a standard deviation of $s_\sigma$ (measured in units of the index).

\subsubsection{Half-normal distribution for $T_{P}$\label{sec:gaussian}}
We explore how a half-normal distribution of $T_{P}$ affects
the derived values of $n$ and SNR.
A half-normal distribution can be a good approximation for data sets where the brightest spectra have a much higher signal strength than the mean.  If $T_{P}$ follows a half-normal
distribution,
\begin{equation}
  T_{P,i} = T_{P,{\rm max}}\exp{\left(-\frac{i^2}{2s_T^2}\right)}\,,
  \label{eq:gauss}
\end{equation}
where $T_{P, {\rm max}}$ is the maximum line height in the dataset and the standard deviation in the distribution of $T_{P}$ is $s_T$ (measured in units of the index).  From Equation~\ref{eq:snr}, the SNR is then
\begin{equation}
\begin{split}
  {\rm SNR} = C T_{P, {\rm max}} \left(\frac{\Delta V}{\Delta\lambda}\right)^{0.5}
  \frac{\sum_{i=1}^n
 w_i\exp{\left(-\frac{i^2}{2s_T^2}\right)}}{\left(\sum_{i=1}^n \sigma_{i}^2 w_i^2 \right)^{0.5}}
\label{eq:snr_gauss}
\end{split}
\end{equation}

To illustrate the basic functional dependencies, we can assume that the noise is uncorrelated, is the same in all spectra to be averaged, and is equal to $\sigma_0$.  In this case, for 
intensity-noise weighting, we have \begin{equation}
  {\rm SNR} = C \frac{T_{P, {\rm max}}}{\sigma_0}
  \left(\frac{\Delta V}{\Delta\lambda}\right)^{0.5} \exp{(s_T)} \left[\sum_{i=1}^n \exp{(-i^2)}\right]^{0.5}\,.
\label{eq:snr_intensity_gauss}
\end{equation}
For noise and uniform weighting, we have
\begin{equation}
  {\rm SNR} = C \frac{T_{P, {\rm max}}}{\sigma_0}
  \left(\frac{\Delta V}{\Delta\lambda}\right)^{0.5}\exp{(0.5s_T)} \left[\sum_{i=1}^n \exp{(-i^2)}\right]^{0.5}\,.
  \label{eq:snr_noise_gauss}
\end{equation}
As can be seen in
Equations~\ref{eq:snr_intensity_gauss} and \ref{eq:snr_noise_gauss}, in the case of constant noise the SNR depends on the
ratio of the maximum line intensity divided by the noise, rather
than the individual value of either quantity.  We use this ratio to parameterize the simulations.

We create 100 simulated peak signal and noise distributions for values of
$T_{P, {\rm max}}/\sigma_0$ of 0.01, 0.05, 0.1, 0.5, 1, 2,
  and 5 in two noise distributions: ``constant'' noise (all spectra have the same noise value) and Gaussian noise with
$s_\sigma = 0.1$ (each spectrum has a noise value drawn randomly from a normal distribution).  All signal distributions have a standard deviation $s_T=1.5$.  For the Gaussian noise trials, we split the
analysis into two categories: 1) the estimation of the signal strength is unaffected by
the noise; and 2) the estimation of the signal strength is modified by the normal distribution of standard deviation $s_\sigma$.
The former case represents the theoretical situation when noise does not affect the estimation of the signal; the latter case is more realistic.  We analyze both cases to determine how noise affects the analysis.

\begin{figure*}
  \includegraphics[width=0.5\textwidth]{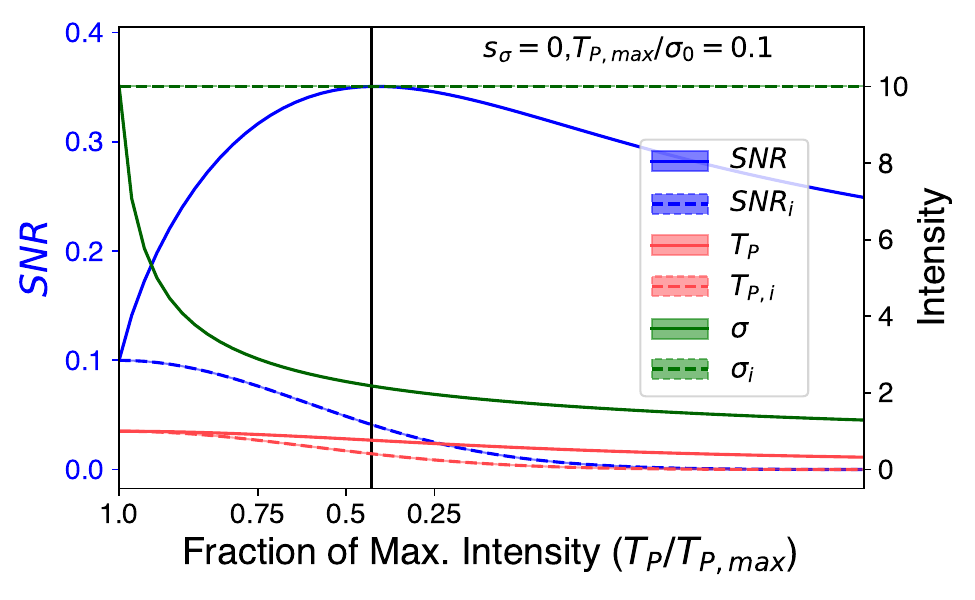}
  \includegraphics[width=0.5\textwidth]{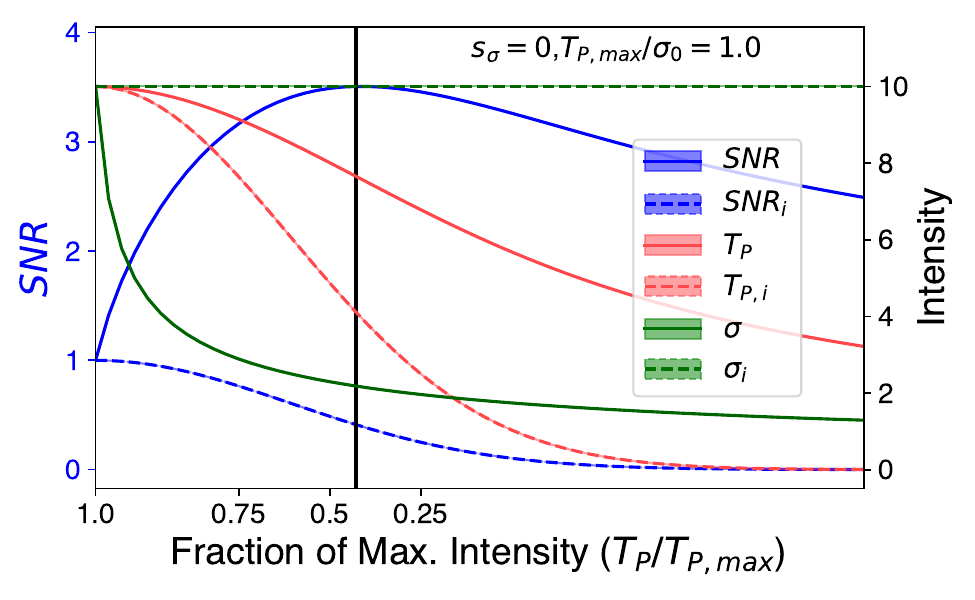}
  \includegraphics[width=0.5\textwidth]{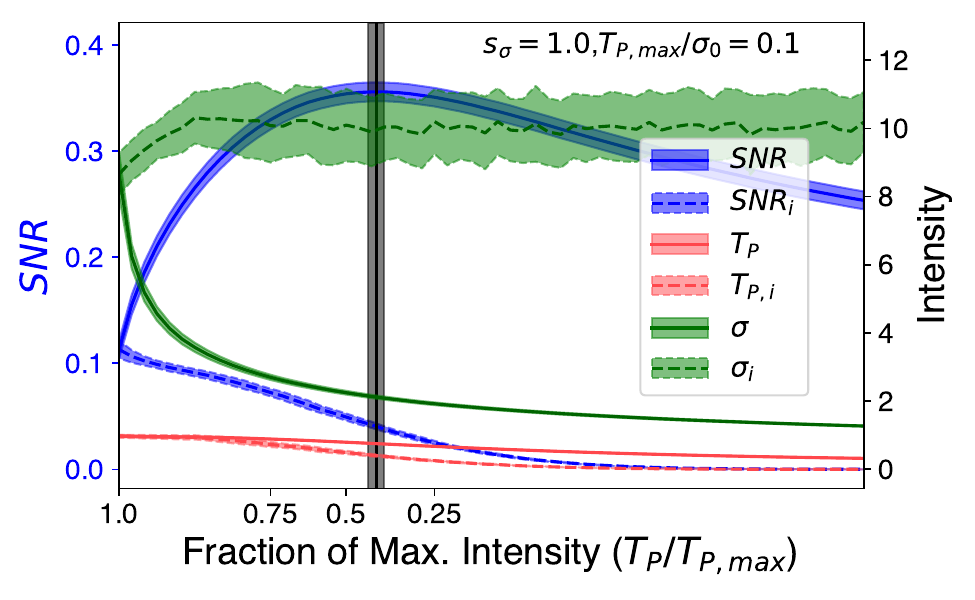}
  \includegraphics[width=0.5\textwidth]{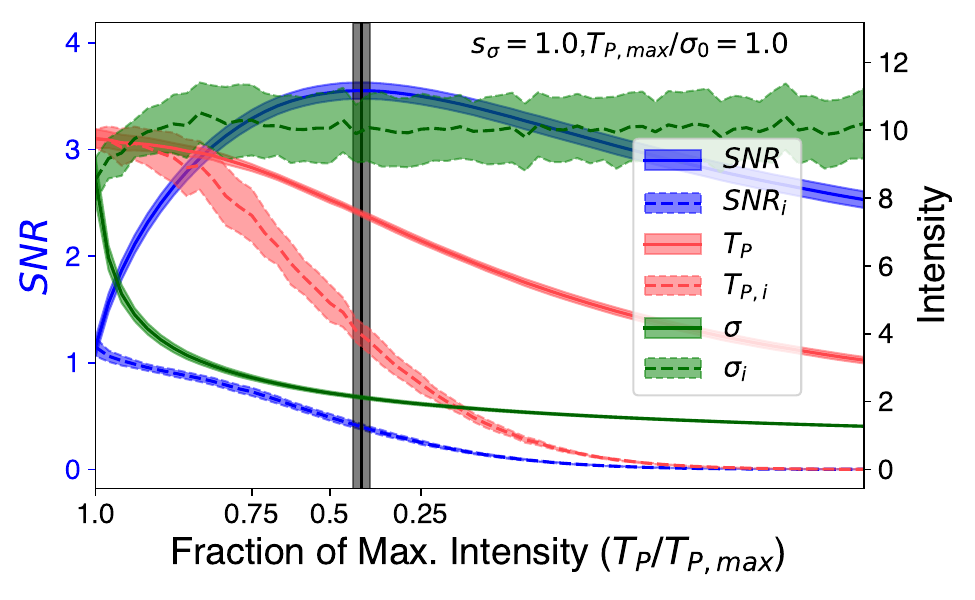}
  \caption{SNR analysis for half-normal-distributed values of
    $T_{P}$, with noise weighting.  The SNR in intensity-noise weighting (not shown) increases without bound whereas uniform weighting (also not shown) produces nearly identical results to those of noise weighting. Panels in the left column show $T_{P,
      max} / \sigma_0=0.1$ and those of the right column show $T_{P, {\rm max}} / \sigma_0 = 1.0$.  The top row of panels has $s_\sigma =
    0$ (constant noise) and the bottom row has $s_\sigma = 0.1 \sigma_0$.
    In
    all panels, solid lines show individual values and dashed lines
    show integrated values. The blue curves show the SNR (and use the left y-axis), the red
    curves show $T_P$, and the green curves show the noise (both use the right y-axis).  The shaded regions in the lower panels show the standard
    deviations from the Monte Carlo simulations. The vertical gray lines indicate the peaks of the SNR distributions; the gray shaded areas show the range within one standard deviation of the SNR distributions.
    \label{fig:sn_gauss}}
\end{figure*}

For each set of distributions, we estimate 
SNR$_{\rm max}$ using Equation~\ref{eq:snr}.  For noise and uniform weighting trials, we additionally compute the signal at the maximum SNR compared to the maximum signal,
$T_{P,i=n}/ T_{P,{\rm max}}$.  We give our results from all three weighting schemes in Table~\ref{tab:results-gauss}
and show the noise-weighting analysis in Figure~\ref{fig:sn_gauss} (uniform weighting produces nearly identical results).

Intensity-noise weighting leads to an increase in the SNR without bound and therefore sets SNR$_{\rm max}$ for any averaging method. For constant-noise
half-normal signal distributions and noise or uniform weighting we find:
\begin{itemize}
  \item SNR$_{\rm max} \simeq 3.5 T_{P, {\rm max}} /
  \sigma_0$;
  \item SNR$_{\rm max}$ is obtained when averaging all spectra
    satisfying $T_{P} \gtrsim 0.4 T_{P, {\rm max}}$;
    \item SNR$_{\rm max}$ is $\sim\!5\%$ less than that from averaging all spectra using intensity-noise weighting, assuming $T_P$ can be reliably estimated.
    \item The SNR can decrease by up to 30\% relative to SNR$_{\rm max}$ when averaging spectra down to $T_P/T_{P, {\rm max}}\simeq 0.01$.
    \item SNR$_{\rm max}$ for noise and uniform weighting is $\sim\!95\%$ that found for intensity-noise weighting.
\end{itemize}
These relationships also hold for the variable noise distributions when the noise and signal strength are uncorrelated.
The above are theoretical best-case scenarios. If noise affects the estimation of the 
signal strength, as it does for actual data, the inability to reliably order the highest SNR
spectra affects the SNR; these effects are larger if the noise is comparable to the signal.

\begin{deluxetable*}{rcccccccc}
  \tablecaption{SNR analysis for half-normal intensity distributions  \label{tab:results-gauss}}
  \tablehead{
    \colhead{} & 
    \colhead{} &
    \colhead{Intensity-noise} &
    \colhead{}  & 
    \multicolumn{2}{c}{Noise} &
    \colhead{}  & 
    \multicolumn{2}{c}{Uniform}\\ \cline{3-3} \cline{5-6} \cline{8-9}
    \colhead{}  & 
    \colhead{$T_{P, {\rm max}} / \sigma_0$} &
    \colhead{SNR$_{\rm max}$} & 
    \colhead{} &
    \colhead{SNR$_{\rm max}$} &
    \colhead{$T_{P, i=n}/T_{P, {\rm max}}$} &
    \colhead{} &
    \colhead{SNR$_{\rm max}$} &
    \colhead{$T_{P, i=n}/T_{P, {\rm max}}$}}
  \startdata
  \multirow{6}{*}{$\sigma_i = \sigma_0$ $\begin{dcases*} \\ \\ \\ \\ \\ \end{dcases*}$}
  & 0.01 & 0.037 && 0.035 & 0.41 && 0.035 & 0.41\\
  & 0.05 & 0.19 && 0.18  & 0.41 && 0.18 & 0.41\\
  & 0.1  & 0.37 && 0.35  & 0.41 && 0.35 & 0.41\\
  & 0.5  & 1.9 && 1.8   & 0.41 && 1.8 & 0.41\\
  & 1.0  & 3.7 && 3.6   & 0.41 && 3.6 & 0.41\\
  & 5.0  & 19 && 18    & 0.41 && 18 & 0.41\\[1em]
  \multirow{6}{*}{$s_\sigma=0.1 \sigma_0$ $\begin{dcases*} \\ \\ \\ \\ \\ \end{dcases*}$}
  & 0.01 & 0.038 &&0.036 & 0.42 && 0.035 & 0.37\\
  & 0.05 & 0.18  &&0.19 & 0.42 && 0.18 & 0.39\\
  & 0.1  & 0.38 && 0.35  & 0.43 && 0.35 & 0.40\\
  & 0.5  & 1.9 && 1.8   & 0.42 && 1.7 & 0.39\\
  & 1.0  & 3.8 && 3.6   & 0.43 && 3.5 & 0.39\\
  & 5.0  & 19 && 18    & 0.42 && 18 & 0.40\\[1em]
  \multirow{6}{*}{$\begin{dcases*} \\ \\ \\ \\ \\ \end{dcases*}$}
  & 0.01 & 0.038 && 0.026  & 0.67 && 0.038 & 0.26\\
  & 0.05 & 0.19 && 0.13  & 0.66 && 0.13 & 0.65\\
  $s_\sigma=0.1 \sigma_0$\,\,\,\,\,\,\, & 0.1  & 0.38 && 0.27  & 0.68 && 0.26 & 0.65\\
  $T_{P, i}$ modified\,\,\,\,\,\,\, & 0.5  & 1.9 && 1.7   & 0.60 && 1.7 & 0.61\\
  & 1.0  & 3.8 && 3.5   & 0.49 && 3.5 & 0.48\\
  & 5.0 & 19  && 18    & 0.42 && 18 & 0.40\enddata
\end{deluxetable*}

\subsubsection{Power Law distribution for $T_{P}$\label{sec:powerlaw}}
We perform a similar analysis assuming a power law distribution for
$T_{P}$:
\begin{equation}
  T_{P, i} = T_{P, {\rm max}} i^{-\alpha}\,,
  \label{eq:pow}
\end{equation}
where the maximum value is $T_{\rm P, {\rm max}}$ and $\alpha$ is the
power law index.  The relevant SNR equation is then
\begin{equation}
  {\rm SNR} = C T_{P, {\rm max}} \left(\frac{\Delta V}{\Delta\lambda}\right)^{0.5}
  \frac{\sum_{i=1}^n
 w_i\ i^{-\alpha}}{\left(\sum_{i=1}^n w_i^2 \sigma_i^2\right)^{0.5}}
\label{eq:snr_powerlaw}
\end{equation}

In the case of constant noise, 
for intensity-noise weighting, we have
\begin{equation}
  {\rm SNR} = C \frac{T_{P, {\rm max}}}{\sigma_0}
  \left(\frac{\Delta V}{\Delta\lambda}\right)^{0.5} \left(\sum_{i=1}^n i^{-2\alpha} \right)^{0.5}\,.
  \label{eq:snr_intensity_powerlaw_const}
\end{equation}
For noise and uniform weighting, we have
\begin{equation}
  {\rm SNR} = C \frac{T_{P, {\rm max}}}{\sigma_0}
  \left(\frac{\Delta V}{\Delta\lambda}\right)^{0.5} \left( \sum_{i=1}^n i^{-\alpha} \right) n^{-0.5}\,.
  \label{eq:snr_noise_powerlaw_const}
\end{equation}
Once again we see that the SNR is linearly proportional to the ratio of the maximum peak to the standard deviation.

\begin{figure*}[!ht]
  \includegraphics[width=0.5\textwidth]{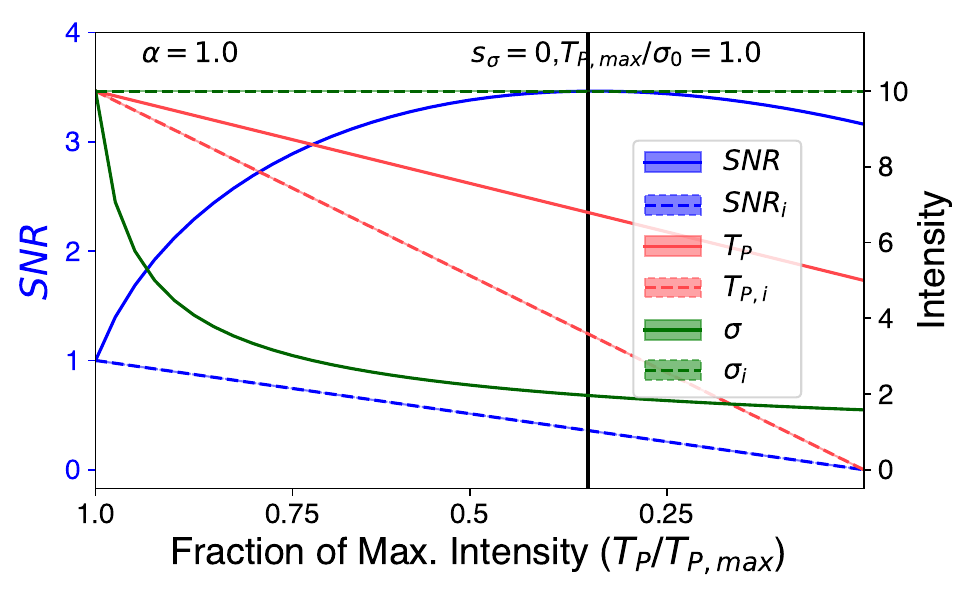}
  \includegraphics[width=0.5\textwidth]{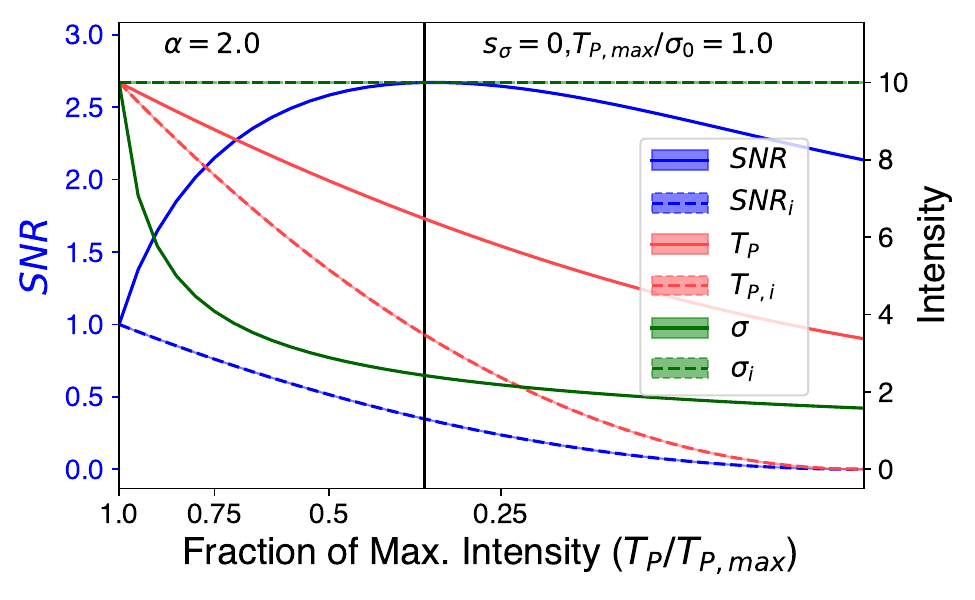}
  \includegraphics[width=0.5\textwidth]{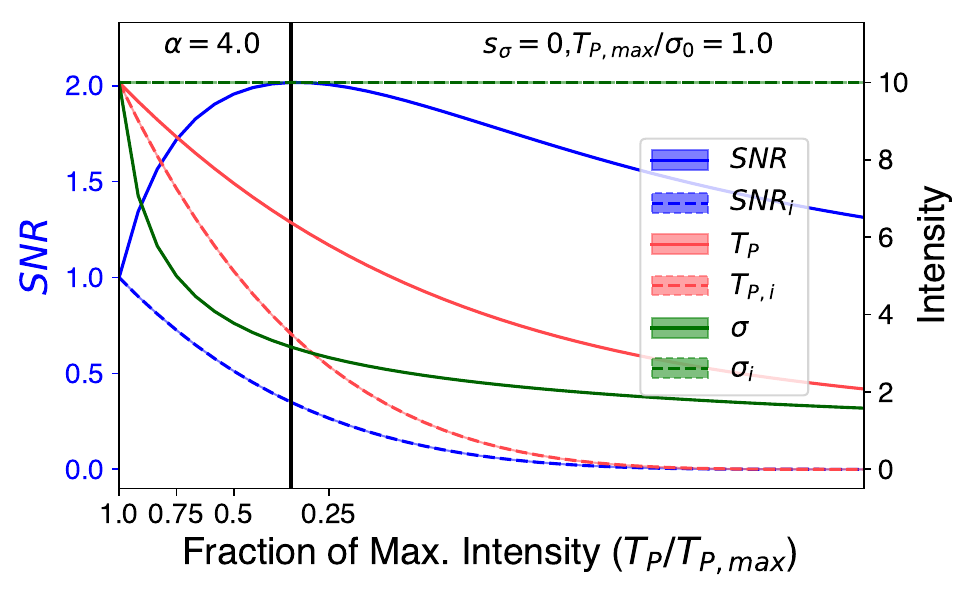}
  \includegraphics[width=0.5\textwidth]{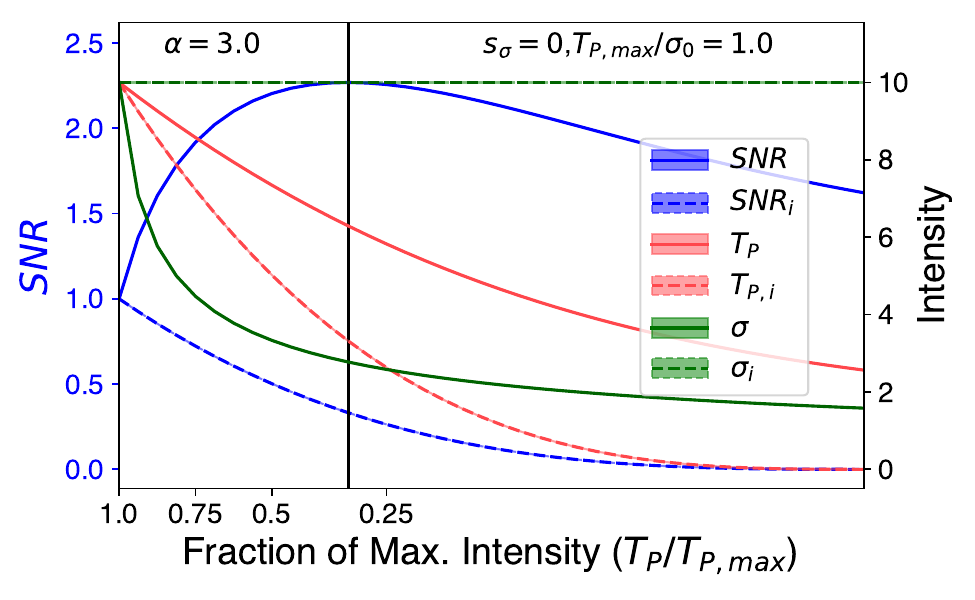}
  \caption{SNR analysis for power law-distributed values
    of $T_{P}$, with noise weighting.  {Uniform weighting (not shown) produces nearly identical results.}. All panels have ${T_P,
      {\rm max}}/\sigma_0 = 1.0$, and clockwise from the top-left panel
    $\alpha$ ranges from 1 to 4 in integer steps.  In all panels,
    solid lines show individual values and dotted lines show
    integrated values. The blue curves show the SNR, the red curves
    show $T_{P}$, and the green curves 
    show the noise.  The vertical gray lines indicates the peaks of the SNR distribution.
    \label{fig:sn_pow}}
\end{figure*}

We investigate the effect of different power law distributions for $\alpha=0.5$ to $4.5$ in increments of 0.5 with ${T_P, {\rm max}}/\sigma_0 = 1.0$, and for
${T_P, {\rm max}}/\sigma_0 = 0.1,0.5,1.0,5.0,10$ and 50 with $\alpha = 2.0$.  We show
these results in Figures~\ref{fig:sn_pow} and in Table~\ref{tab:pow-results}.  We do not consider
variable noise (and so set $s_\sigma=0$), which we assume has a minor effect, as it does for the
half-normal distributions.  For a constant-noise power law signal
distribution with noise or uniform weighting, we find:
\begin{itemize}
  \item SNR$_{\rm max}$ decreases with increasing power law index
    and decreasing values of $T_{P, {\rm max}}/\sigma_0$;
  \item SNR$_{\rm max}$ is obtained when averaging all spectra
    satisfying $T_{P} \gtrsim 0.35 {T_P, {\rm max}}$ for the values of $\alpha$ and ${T_P, {\rm max}}/\sigma_0$ investigated;
    \item As for the half-normal signal distribution, SNR$_{\rm max}$ is $\sim\!5\%$ less than that from averaging all spectra using intensity-noise weighting.
    \item The SNR can decrease by up to 30\% relative to SNR$_{\rm max}$ when averaging spectra down to $T_P/T_{P, {\rm max}}\simeq 0.01$.
    \item SNR$_{\rm max}$ for noise and uniform weighting is $\sim\!95\%$ that found for intensity-noise weighting.
    \end{itemize}


\begin{deluxetable*}{ccccccccc}[!ht]
  \label{tab:pow-results}
  \tablecaption{SNR analysis for power law intensity distributions}
  \tablehead{
    \colhead{}  & 
    \colhead{}  & 
    \colhead{Intensity-noise} &
    \colhead{}  & 
    \multicolumn{2}{c}{Noise} &
    \colhead{}  & 
    \multicolumn{2}{c}{Uniform} \\ \cline{3-3} \cline{5-6} \cline{8-9}
    \colhead{$T_{P, max} / \sigma_0$} & \colhead{$\alpha$} & 
    \colhead{SNR$_{\rm max}$} & 
    \colhead{} &
    \colhead{SNR$_{\rm max}$} &
    \colhead{$T_{P, i=n}/T_{P, {\rm max}}$} &
    \colhead{} &
    \colhead{SNR$_{\rm max}$} &
    \colhead{$T_{P, i=n}/T_{P, {\rm max}}$}}
  \startdata
  1.0 & 0.5 & 4.5 && 4.3    & 0.39 && 4.3    & 0.39\\
  1.0 & 1.0 & 3.7 && 3.5    & 0.36 && 3.5    & 0.36\\
  1.0 & 1.5 & 3.2 && 3.0    & 0.34 && 3.0    & 0.34\\
  1.0 & 2.0 & 2.9 && 2.7    & 0.35 && 2.7    & 0.35\\
  1.0 & 2.5 & 2.6 && 2.4    & 0.33 && 2.4    & 0.33\\
  1.0 & 3.0 & 2.5 && 2.3    & 0.33 && 2.3    & 0.33\\
  1.0 & 3.5 & 2.3 && 2.1    & 0.35 && 2.1    & 0.35\\
  1.0 & 4.0 & 2.2 && 2.0    & 0.35 && 2.0    & 0.35\\
  1.0 & 4.5 & 2.1 && 1.9    & 0.36 && 1.9    & 0.36\\[1em]
  0.1 & 2.0 & 0.029 && 0.027  & 0.35 && 0.027  & 0.35\\
  0.5 & 2.0 & 0.14 && 0.13   & 0.35 && 0.13   & 0.35\\
  1.0 & 2.0 & 0.29 && 0.27   & 0.35 && 0.27   & 0.35\\
  5.0 & 2.0 & 1.4 && 1.3    & 0.35 && 1.3    & 0.35\\
  10  & 2.0 & 2.9 && 2.7    & 0.35 && 2.7    & 0.35\\
  50  & 2.0 &  14 && 13     & 0.35 && 13     & 0.35\\
  \enddata
\end{deluxetable*}

\section{Application to GDIGS data\label{sec:real_data}}
The GBT Diffuse Ionized Gas (GDIGS) survey \citep{anderson21} traced the radio recombination line (RRL) emission across the inner Galaxy, over $-5\degree<\ell<32\degree$, $\absb < 0.5\degree$.  The data were collected using the C-band receiver on the Green Bank Telescope (GBT) in total power mode.  Within the 4--8\,\ghz\ bandpass, GDIGS tuned to 15 usable hydrogen RRLs and averaged their signals to produce the reduced \hna\ data set. The reduced data have a spatial resolution of 2\arcmper65, a spaxel size of $30\arcsec$, and a  spectral resolution of $0.5\,\kms$. 
The rms spectral noise per spaxel is $\sim\!10$\,mK.

We test the above spectral averaging methods using GDIGS data to constrain the ionic $\rm \he4^+/\,H^+$ abundance ratio by number, $y^+$.  Measurements of elemental abundances
provide key constraints for our understanding of Galactic chemical
evolution.  We define $y^+$ as
\begin{equation}
y^+=\frac{T_{\rm P, He} \Delta V_{\rm He}}{T_{\rm P, H} \Delta V_{\rm H}}
\label{eq:yplus}
\end{equation}
where $T_{\rm P}$ is the peak line intensity and $\Delta V$ is the FWHM
line width. The uncertainty on $y^+$ is therefore
\begin{equation}
\begin{split}
    \sigma_{y^+} = y^+ \biggl[\left(\frac{\sigma_{T_{\rm P,He}}}{T_{\rm P,He}}\right)^2 + \left(\frac{\sigma_{\Delta V_{\rm He}}}{\Delta V_{\rm He}}\right)^2 +\\
\left(\frac{\sigma_{T_{\rm P,H}}}{T_{\rm P,H}} \right)^2 + \left(\frac{\sigma_{\Delta V_{\rm H}}}{\Delta V_{\rm H}} \right)^2 \biggr]^{0.5}\,,
\end{split}
\label{eq:e_yplus}
\end{equation}
where $\sigma$ denotes parameter uncertainties.
If the
source is optically thin, $y^+$
measures the $\rm \he4^+/\,H^+$ abundance ratio directly.  

Because the mass of helium is greater than that of hydrogen, its RRL velocity is shifted by $\sim -122\,\kms$ from that of hydrogen.  Both lines therefore fall within the same GDIGS bandpass and are subject to the same systematic effects.


To spectrally average GDIGS RRL data using intensity-noise weighting, we:
\begin{itemize}
    \item align the spectra in velocity using the velocity centroids from the Automatic Gaussian Decomposition (AGD) results described in \citet{anderson21} \citep[which in turn use the code from][]{ riener19};
    \item average all spectra;
    \item remove a fifth-order polynomial baseline and determine the SNR in the average spectrum using Gaussian fits to the hydrogen RRLs.
\end{itemize}
To spectrally average GDIGS RRL data using noise or uniform weighting, we:
\begin{itemize}
    \item determine the SNR and peak intensity for each spaxel using the results from the AGD analysis;
    \item align the spectra in velocity using the velocity centroids from the AGD analysis;
    \item average spectra, starting with the highest SNR spectrum;
    \item remove a fifth-order polynomial baseline from line-free portions of the spectrum;
    \item determine the SNR in the average spectrum using Gaussian fits to the hydrogen RRLs;
    \item and cease averaging when the average spectrum SNR stops increasing, with a buffer of 100 spectra (once a peak in SNR is reached, continue averaging the next 100 to determine if the SNR peak is local).
\end{itemize}
For all weighting schemes, we only use spaxels fit by a single Gaussian component in the AGD.  We determine $y^+$ for all average spectra by fitting the helium line using velocities from $-135\,\kms$ to $-110\,\kms$ and the hydrogen line using velocities from $-20\,\kms$ to $+20\,\kms$.



\begin{figure}
    \centering
    \includegraphics[width=3.5in]{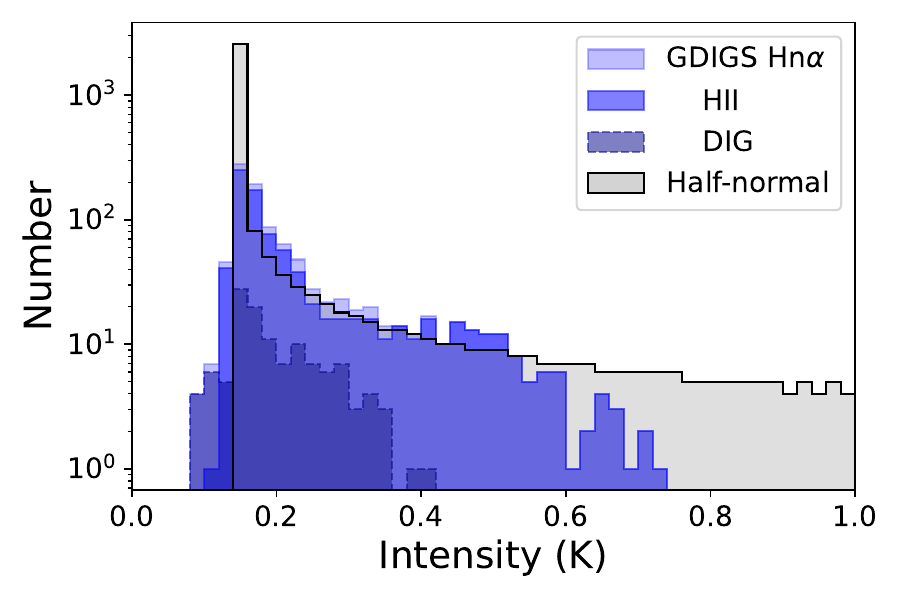}
    \caption{GDIGS data towards W43. 
 Shown are the 1000 highest intensity fitted line height values ($T_P$), separated into those that are spatially coincident with \hii\ regions (``\hii'') and those that are not (``DIG''; diffuse ionized gas).  The ``\hii'' spectra are more numerous in this field at all intensities studied, and all intensities $T_P>0.4\,\K$ are cospatial with \hii\ regions.  A half-normal distribution fits the lower intensity values well, but drastically over-predicts the high values, indicating that the intensity distribution is more complicated than the simple models considered here.}
    \label{fig:gdigs_intensity}
\end{figure}

\begin{figure*}
    \centering
    \includegraphics[width=5in]{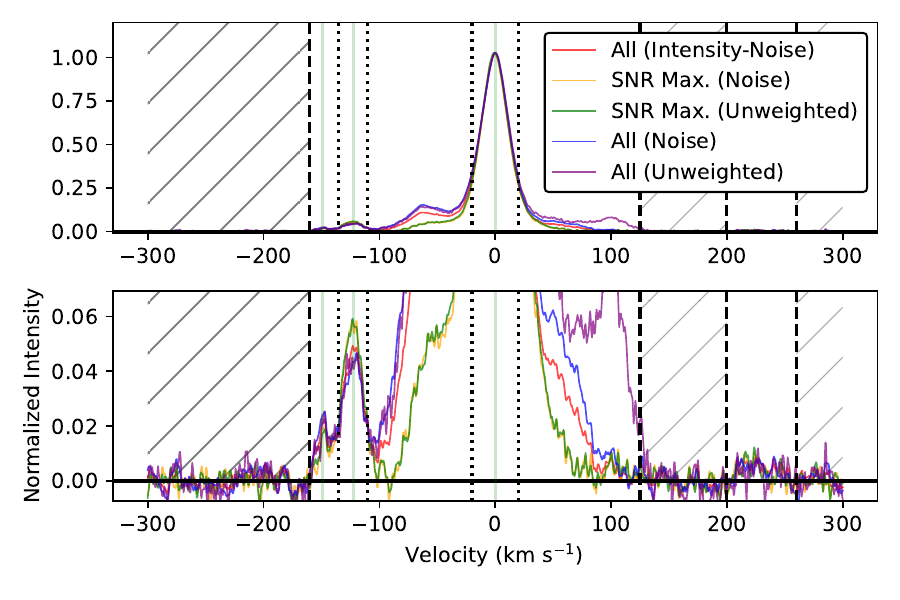}
    \caption{Spectra created using the averaging and weighting methods discussed in the text. The data are the same in both panels, but we adjust the y-axis in the lower panel to better show the helium line centered at $-122\,\kms$. Line-free regions used to fit the baseline are shaded in diagonal gray lines. Dotted black lines show the boundaries used in the fit. Light green vertical lines show the expected velocities of the hydrogen, helium, and carbon RRLs.}
    \label{fig:gdigs}
\end{figure*}

We perform this analysis on the GDIGS \hna\ data in a $100\arcmin \times 60\arcmin$ zone centered on the massive star forming region W43 that was first analyzed in \citet{luisi20}.  The GDIGS data of this zone has 24,000 spectra, of which 19,849 are fit in the AGD with a single hydrogen line.  This zone has numerous \hii\ regions and also diffuse ionized gas \citep[see][]{luisi20}.  We show the distribution of AGD-derived peak line intensities in Figure~\ref{fig:gdigs_intensity} for the 1000 highest-intensity values in the field.  We also separate this distribution into those derived from spaxels falling within \hii\ regions defined by the WISE Catalog of Galactic \hii\ Regions \citep[][hereafter the ``WISE Catalog'']{anderson14}, and those that do not fall within \hii\ regions.  The peak line intensities approximately follow a half-normal distribution of 4000 values with $s_T=400$, $T_{P, {\rm max}}=10\,\K$, and that is scaled so the minimum value is 0.14\,\K (a power law with $\alpha = 4$ also fits fairly well).  The exception to this good fit is at intensities $\gtrsim0.6\,\K$ where the model over-predicts the data.  Thus, the signal distribution is more complicated than the simulated distributions considered here.  Most of the values, and a greater fraction of the high-intensity values, are associated with \hii\ regions. 

We create five different average spectra and compute $y^+$ for each: intensity-noise weighting all spectra, noise weighting with SNR maximization, uniform weighting with SNR maximization, noise weighting all spectra and uniform weighting all spectra. 
The noise and uniform SNR maximizations use 569 and 1102 of the $\sim\!20,\!000$ spectra, respectively, corresponding to approximate intensity values of $T_P>0.15\,\K$ and $T_P >0.14\,\K$.

We show the five average spectra in Figure~\ref{fig:gdigs}.  Each spectrum is independently normalized. All five spectra have the same basic shape, although the SNR maximization spectra have the smallest deviations from a single Gaussian line.
 In Table~\ref{tab:yplus} we summarize the H and He line height ($T_P$) and FWHM line width ($\Delta V$) for the H and He RRLs, as well as their $1\sigma$ fit uncertainties, $y^+$ (Equation~\ref{eq:yplus}) and its $1\sigma$ uncertainty (Equation~\ref{eq:e_yplus}), and the spectral rms $\sigma$.  The derived values of $y^+$ differ depending on the averaging method and the weighting scheme. 
 Differences in $y^+$ are not accounted for by the uncertainties in $\sigma_{y^+}$.  As expected, uniformly weighting all spectra results in the largest rms spectral noise; the other spectral noise values are similar.

\begin{deluxetable*}{lcccccccccccc}
  \tablecaption{Analysis of GDIGS data\tablenotemark{a}\label{tab:yplus}}
  \tablehead{
& \multicolumn{4}{c}{H} && \multicolumn{4}{c}{He} &&&\\ \cline{2-5}\cline{7-10} 
         Weighting &  
         $T_P$ & $\sigma_{T_P}$ & 
         $\Delta V$ & $\sigma_{\Delta V}$ &&
         $T_{P, {\rm max}}$ & $\sigma_{T_P}$ & 
         $\Delta V$ & $\sigma_{\Delta V}$ &
         $y^+$ & $\sigma_{y^+}$ & $\sigma$\\
         &\K&\K&\kms&\kms&&\K&\K&\kms&\kms&&&\K
         }
\startdata
Intensity-Noise (All) &  1.00  &  0.00329  &  28.4  &  0.133  &&  0.0482  &  0.000677  &  21.6  &  0.532  &  0.037  &  0.0011  &  0.0028 \\
SNR Max (Noise) &  1.00  &  0.00287  &  27.9  &  0.112  &&  0.0578  &  0.000989  &  21.0  &  0.611  &  0.044  &  0.0015  &  0.0039 \\
SNR Max (Unweighted) &  1.00  &  0.00445  &  29.0  &  0.166  &&  0.0571  &  0.000675  &  21.5  &  0.445  &  0.042  &  0.0011  &  0.0037 \\
Noise (All) &  1.00  &  0.00364  &  29.6  &  0.160  &&  0.0438  &  0.000733  &  23.1  &  0.739  &  0.034  &  0.0013  &  0.0036 \\
Unweighted (All) &  1.00  &  0.00383  &  29.8  &  0.170  &&  0.0460  &  0.00153\phn  &  22.2  &  1.34\phn  &  0.034  &  0.0024  &  0.0076
\enddata
\tablenotetext{a}{All intensity values are normalized such that the hydrogen line intensity has a value of 1.00.}
\end{deluxetable*}

\section{Discussion and Summary}
In this paper we explored methods for averaging spectra.  
Intensity-noise weighting leads to the highest possible SNR.  For noise and uniform weighting, averaging the $35-45\%$ highest intensity individual spectra (assuming similar noise characteristics for each) results in the maximum SNR average spectrum, in agreement with the results of \citet{zhang99}. This average spectrum created from the $35-45\%$ highest intensity individual spectra has $\sim\!95\%$ the SNR of the intensity-noise weighted average spectrum. Our results are largely independent of the intensity distribution; other peaked signal distributions should have similar results. 


We apply our averaging methods to Green Bank Telescope (GBT) Diffuse Ionized Gas (GDIGS) \hna\ data \citep{anderson21} to determine
the ionic abundance ratio, $y^+$.
The different averaging methods give values of $y^+$ that differ by $\sim\!25\%$.

Differences in the derived values of $y^+$ can be explained by which locations are weighted more heavily during averaging. Intensity-noise weighting obviously preferences the spectra with the highest peak intensity.  For GDIGS, the highest intensities are found toward discrete \hii\ regions; the highest intensity diffuse regions are found just outside of the discrete \hii\ regions \citep[see][]{luisi20}.  Noise weighting preferences the spectra with the lowest noise, whereas uniform weighting weights all spectra evenly.  Since \hii\ regions have bright radio continuum emission, noise weighting can preference the diffuse regions.  The SNR maximization method only averages the highest SNR spectra, which means that only the brightest regions may appear in the average, regardless of their noise levels.

That the value derived for $y^+$ depends on the weighting scheme employed indicates that there are differences in $y^+$ in the GDIGS field studied; if $y^+$ were invariant, all averaging techniques would produce the same result.  This piece of evidence is not as apparent without averaging, as the He RRL signal that goes into the $y^+$ computation is weak and can only be seen in a fraction of the GDIGS spectra.  We caution that studies of $y^+$ that include a range of intensity values (i.e., from both \hii\ regions and from diffuse ionized gas, as in our example) will be biased depending on the weighting scheme. In future research with the GDIGS data, we will investigate and model $y^+$ over the survey area with these considerations in mind.


The SNR maximizing procedure allows for the creation of more sensitive spectra, and therefore a more accurate determination of $y^+$, but the derived $y^+$ values in all average GDIGS spectra are low relative to those found previously for Galactic \hii\ regions.  For comparison, an analysis of the 80 high-quality RRL spectra toward \hii\ regions in \citet{quireza06a} by \citet{wenger13} found
$y^+ = 0.075\pm0.024$. \citet{wenger13} found $y^+ = 0.068\pm0.023$ in a sample of 54 high-quality RRL spectra towards Galactic \hii\ regions.  For the \hii\ region W43, which is in the studied field, $y^+ = 0.068 \pm 0.0052$ \citet{bania97, bania07}.
It may be that the inclusion of the diffuse ionized gas outside of \hii\ regions has caused the discrepancy with values derived for \hii\ regions; we will investigate the cause of the low $y^+$ values in a subsequent paper.

\acknowledgments
We thank the anonymous referee, whose thoughtful comments greatly improved the clarity of this manuscript.
This work is supported by NSF grant AST1516021 to L.D.A. The Green Bank Observatory is a facility of the National Science Foundation operated under cooperative agreement by Associated Universities, Inc.
T.V.W. is supported by a National Science Foundation Astronomy and
Astrophysics Postdoctoral Fellowship under award AST-2202340.


\bibliographystyle{aasjournal.bst}
\bibliography{ref.bib} 

\begin{thebibliography}{}
\expandafter\ifx\csname natexlab\endcsname\relax\def\natexlab#1{#1}\fi
\providecommand{\url}[1]{\href{#1}{#1}}

\bibitem[{{Anderson} {et~al.}(2014){Anderson}, {Bania}, {Balser}, {Cunningham},
  {Wenger}, {Johnstone}, \& {Armentrout}}]{anderson14}
{Anderson}, L.~D., {Bania}, T.~M., {Balser}, D.~S., {et~al.} 2014, \apjs, 212,
  1

\bibitem[{{Anderson} {et~al.}(2021){Anderson}, {Luisi}, {Liu}, {Wenger},
  {Balser}, {Bania}, {Haffner}, {Linville}, \& {Mascoop}}]{anderson21}
{Anderson}, L.~D., {Luisi}, M., {Liu}, B., {et~al.} 2021, \apjs, 254, 28

\bibitem[{{Bania} {et~al.}(2007){Bania}, {Balser}, {Rood}, {Wilson}, \&
  {LaRocque}}]{bania07}
{Bania}, T.~M., {Balser}, D.~S., {Rood}, R.~T., {Wilson}, T.~L., \& {LaRocque},
  J.~M. 2007, \apj, 664, 915

\bibitem[{{Bania} {et~al.}(1997){Bania}, {Balser}, {Rood}, {Wilson}, \&
  {Wilson}}]{bania97}
{Bania}, T.~M., {Balser}, D.~S., {Rood}, R.~T., {Wilson}, T.~L., \& {Wilson},
  T.~J. 1997, \apjs, 113, 353

\bibitem[{{Cappellari} \& {Copin}(2003)}]{cappellari03}
{Cappellari}, M., \& {Copin}, Y. 2003, \mnras, 342, 345

\bibitem[{{Diehl} \& {Statler}(2006)}]{diehl06}
{Diehl}, S., \& {Statler}, T.~S. 2006, \mnras, 368, 497

\bibitem[{{Lenz} \& {Ayres}(1992)}]{lenz92}
{Lenz}, D.~D., \& {Ayres}, T.~R. 1992, \pasp, 104, 1104

\bibitem[{{Luisi} {et~al.}(2020){Luisi}, {Anderson}, {Liu}, {Balser}, {Bania},
  {Wenger}, \& {Haffner}}]{luisi20}
{Luisi}, M., {Anderson}, L.~D., {Liu}, B., {et~al.} 2020, \apj, 889, 96

\bibitem[{{Quireza} {et~al.}(2006){Quireza}, {Rood}, {Balser}, \&
  {Bania}}]{quireza06a}
{Quireza}, C., {Rood}, R.~T., {Balser}, D.~S., \& {Bania}, T.~M. 2006, \apjs,
  165, 338

\bibitem[{{Riener} {et~al.}(2019){Riener}, {Kainulainen}, {Henshaw}, {Orkisz},
  {Murray}, \& {Beuther}}]{riener19}
{Riener}, M., {Kainulainen}, J., {Henshaw}, J.~D., {et~al.} 2019, \aap, 628,
  A78

\bibitem[{{Rosales-Ortega} {et~al.}(2012){Rosales-Ortega}, {Arribas}, \&
  {Colina}}]{rosales-ortega12}
{Rosales-Ortega}, F.~F., {Arribas}, S., \& {Colina}, L. 2012, \aap, 539, A73

\bibitem[{{Unser} \& {Eden}(1990)}]{unser90}
{Unser}, M., \& {Eden}, M. 1990, IEEE Transactions on Acoustics, Speech, and
  Signal Processing, 38, 890

\bibitem[{{Wenger} {et~al.}(2013){Wenger}, {Bania}, {Balser}, \&
  {Anderson}}]{wenger13}
{Wenger}, T.~V., {Bania}, T.~M., {Balser}, D.~S., \& {Anderson}, L.~D. 2013,
  \apj, 764, 34

\bibitem[{{Zhang} \& {McElvain}(1999)}]{zhang99}
{Zhang}, Z., \& {McElvain}, J.~S. 1999, Analytical Chemistry, 71, 39

\end{thebibliography}

\end{document}